\documentclass[conference,a4paper]{APSIPA2020}
\usepackage{multirow}
\usepackage{graphicx}%[dvips]
\usepackage{amsmath}
\usepackage[psamsfonts]{amssymb}
\usepackage{amsxtra}
\usepackage{threeparttable}

\usepackage{cite,balance}
\usepackage{epsfig,amssymb,amsmath,multirow}

\usepackage{color}
\usepackage{hyperref}
\usepackage{makecell}
\usepackage{balance}
\usepackage{amsmath}
\usepackage{xcolor,colortbl}

\hypersetup{
   unicode=false,          % non-Latin characters in Acrobat’s bookmarks
   pdftoolbar=true,        % show Acrobat’s toolbar?
   pdfmenubar=true,        % show Acrobat’s menu?
   pdffitwindow=false,     % window fit to page when opened
   pdfstartview={FitH},    % fits the width of the page to the window
   pdftitle={My title},    % title
   pdfauthor={Author},     % author
   pdfsubject={Subject},   % subject of the document
   pdfcreator={Creator},   % creator of the document
   pdfproducer={Producer}, % producer of the document
   pdfkeywords={keyword1} {key2} {key3}, % list of keywords
   pdfnewwindow=true,      % links in new window
   colorlinks=true,       % false: boxed links; true: colored links
   linkcolor=blue,          % color of internal links
   citecolor=blue,        % color of links to bibliography
   filecolor=magenta,      % color of file links
   urlcolor=blue           % color of external links
}

\begin{document}

\title{Emotion Invariant Speaker Embeddings for Speaker Identification with Emotional Speech}

\author{%
\authorblockN{%
Biswajit Dev Sarma\authorrefmark{1} and Rohan Kumar Das\authorrefmark{2}
}
%
% \authorblockA{%
\authorrefmark{1}Indian Institute of Technology Guwahati, Guwahati, India\\
\authorrefmark{2}Department of Electrical and Computer Engineering, National University of Singapore, Singapore\\
% Tsinghua University, Beijing, China \\
% E-mail: fzheng@tsinghua.edu.cn  Tel/Fax: +86-10-XXXXXXXX}
%
% \authorblockA{%
% \authorrefmark{2}
% Department of Electrical and Computer Engineering,\\
% National University of Singapore, Singapore\\
E-mail: biswajit.devsarma@gmail.com, rohankd@nus.edu.sg%}
}

\maketitle
\thispagestyle{empty}

\begin{abstract}

Emotional state of a speaker is found to have significant effect in speech production, which can deviate speech from that arising from neutral state. This makes identifying speakers with different emotions a challenging task as generally the speaker models are trained using neutral speech. In this work, we propose to overcome this problem by creation of emotion invariant speaker embedding. We learn an extractor network that maps the test embeddings with different emotions obtained using i-vector based system to an emotion invariant space. The resultant test embeddings thus become emotion invariant and thereby compensate the mismatch between various emotional states. The studies are conducted using four different emotion classes from IEMOCAP database. We obtain an absolute improvement of 2.6\% in accuracy for speaker identification studies using emotion invariant speaker embedding against average speaker model based framework with different emotions.

\end{abstract}

\section{Introduction}
\label{sec:intro}

Identifying speakers with their unique traits such as voice is often affected by both external and internal factors~\cite{Tomi,hansen_review,wu_2006,Mahesh_icassp2019}. The external factors include environmental noise, channel/session effects and sensor mismatch~\cite{Mclaren+2016_odyssey,Das2015,rkd_is_2016,Alam2018}. On the contrary, the internal factors refer to the health and emotional state of speakers~\cite{is2012_speaker_trait,is2011_speaker_states}. Most of the studies consider neutral speech to recognize speakers and the performance of such systems deviates in presence of external as well as internal factors~\cite{hasan_is2013,Das_durMod, APSIPA_realism,ffsvc2020,TASLP_Alku2016,VESTMAN2018}. 

The internal source mismatch based studies have gained interest of the community over the time as they are mostly related to human computer interactions and criminal investigations, where there is not much control over the mode of speech~\cite{Hansen2011_wshp,whisper_effect,TASLP_Alku2016,VESTMAN2018,rkd_ICASSP_whisper}. In addition, presence of emotion in speech is very common and natural, which cannot be avoided in practical scenarios~\cite{SCHERER_paper}. A study~\cite{SMM_kovid2018} shows that there is more than 37\% of emotional speech present in the data collected from realistic environment. This showcases the need to investigate speaker identification using emotional speech~\cite{emo_study2011}.

% However, there are applications where emotional speech may be preferred. Such applications are mostly related human computer interactions and criminal investigations, where there is not much control over the mode of speech. Moreover, presence of emotion in speech is very common and natural and can't be avoided in any practical situation. We investigated realistic emotional data in our previous work~\cite{SMM_kovid2018}. The statistics in Table 1 of \cite{SMM_kovid2018} clearly shows that there is more than 37\% of emotional speech in the data collected from realistic environment.

%There are many works reported in the literature related to the field of speaker recognition with emotional speech. 

%The majority of prior studies to recognize speakers with emotional speech rely on feature and score normalization techniques.

The earlier attempts in speaker identification with emotional speech include normalization and transformation of features from emotional speech to that of neutral speech~\cite{Koolagudi_2012,KSRao_emo_si}. The mel frequency cepstral coefficient (MFCC) features computed from the emotional speech using auto associative neural network are transformed to be used in a Gaussian mixture model (GMM) based speaker recognition system. Another work studied an emotion dependent score normalization technique for speaker verification~\cite{wu_2006}. The GMM-universal background model (UBM) based speaker verification system performance was reported to be improved for emotional speech after score normalization. The authors of~\cite{chen_2013} performed speaker recognition using model space migration through translated learning. In~\cite{Shahin-2009}, a two-staged approach with hidden Markov model (HMM) was proposed for emotion dependant speaker verification, and later effectiveness of suprasegmental HMMs was shown in~\cite{SHAHIN_paper}. Again, the authors of~\cite{Dongdong-2005} converted the emotional speech to neutral for speaker recognition studies.  

%In another approach, emotion state was converted for speaker recognition and speaker models were generated based on the converted speech \cite{Dongdong-2005}.   However, most of these works are not reported with the state-of-the-art factor analysis or any speaker embedding systems. The i-vector model space used for improving emotional speaker recognition is reported in~\cite{mansour-2016}. In another study, i-vector based speaker recognition reliability was predicted by classifying the speech into broad emotion classes such as arousal and valence \cite{Parthasarathy-2017}. 

Most of the prior works to recognize speakers with emotional speech rely on waveform or feature conversion and score normalization techniques. Further, they did not consider advanced techniques such as i-vectors and x-vectors for speaker modeling~\cite{Dehak2011,xvectors}. A  few studies investigated speaker recognition with emotional speech using i-vectors~\cite{mansour-2016,emo_study_icassp17}. Similarly, the authors of~\cite{Parthasarathy-2017} predicted the i-vector based speaker recognition by classifying the speech into broad emotion classes such as arousal and valence. However, these works with latest systems did not focus much on improving the performance of identifying speakers with emotional speech. This shows the requirement of having robust methods for compensating such mismatch with latest systems, which motivated our current work.

%A few studies i-vector based systems for emotional speech based speaker recognition for initial investigations~\cite{mansour-2016} and reliability predictions for emotion classes~\cite{Parthasarathy-2017}.

% All these works didn't use the modern i-vector based speaker recognition system. In [], the problem of emotional speaker recognition was presented in  i-vector framework. Further, in [] the authors have used the i-vector space model for improving the emotional speaker recognition. 

%We believe characterizing speakers with emotional speech using advanced methods is important as well as making them 

The speaker embedding representation such as i-vectors and x-vectors are used widely for speaker recognition studies~\cite{Dehak2011,xvectors}. Such embeddings capture dominant speaker information that are robust to various external factors like channel/session variation and background noise. We believe it is more preferable to compensate the mismatch due to internal factors such as emotions at the embedding level, rather than at feature or waveform level as that may affect the performance due to external factors. In this regard, we propose an emotion invariant speaker embedding that is independent from emotional effect. These are extracted by using an extractor that is learned using examples obtained through data augmentation at the speaker embedding level. We use i-vector based modeling as a reference system to obtain the speaker embeddings in this study~\cite{Dehak2011}. The extracted i-vectors from different emotions are transformed into an emotion invariant space to normalize the emotion specific information present in them. 

The studies in this work are conducted using IEMOCAP database, which is a standard corpus for emotional speech research~\cite{iemocap}. Four emotion classes, namely, happiness, anger, sadness and neutral are considered for the study. We note that the background models are learned using neutral speech of a large number of speakers with standard well known datasets for speaker recognition. The contribution of this work lies in proposal of emotion invariant speaker embedding for identifying speakers with emotional speech.

\section{Emotion Invariant Speaker Embedding}
\label{secii}

This section details the proposed emotion invariant speaker embedding for speaker identification. We discuss the i-vector based system, emotion invariant extractor and the framework for speaker identification in the following subsections.

%This section details our proposed emotion invariant speaker embedding for speaker identification with emotional speech. The baseline system for speaker identification is based on widely popular i-vector approach~\cite{Dehak2011}. We consider the i-vectors of different emotions from the speaker set to learn the emotion invariant speaker embedding. Next, we discuss the details of the modules of the proposed system.  

\subsection{i-vector based Speaker Representation}

An i-vector is a compact representation for an utterance, which is derived by a factor analysis approach~\cite{Dehak2011}. In this kind of system, the GMM mean supervector of an utterance is projected into a lower dimensional space by a transformation matrix. This transformation matrix, in short T-matrix is learned with a large amount of background speaker population that captures different variabilities present in the speech signal. The low dimensional representation of an utterance thus obtained is referred to as i-vector. However, as it contains channel/session information, some compensation techniques are applied to nullify those unwanted information. In this work, we have chosen linear discriminant analysis (LDA)~\cite{duda_hart} and within class covariance normalization (WCCN)~\cite{Hatch2006} for such compensation on top of i-vectors as given in~\cite{Dehak2011}.

%It started gaining popularity in the current decade. Although recent advances in speaker recognition has lead to systems like recent x-vectors~\cite{xvectors}, the i-vector based system is still popular in the community. In this work, we use i-vectors as a baseline system for the studies. 

%In this kind of system, the GMM mean supervector of an utterance is projected into a lower dimensional space by a transformation matrix. This transformation matrix, in short T-matrix is learned with a large amount of background speaker population that captures different variabilities present in the speech signal. The low dimensional representation of an utterance thus obtained is referred to as i-vector. However, as such compact low dimensional vector posses channel/session information, some compensation techniques are applied on them to nullify those unwanted information. In this work, we have chosen linear discriminant analysis (LDA)~\cite{duda_hart} and within class covariance normalization (WCCN)~\cite{Hatch2006} for such compensation on top of i-vectors as given in~\cite{Dehak2011}. 

\subsection{Emotion Invariant Extractor}

\begin{figure}[t!]
\centering
\vspace{1cm}
\includegraphics[width=0.27\textwidth]{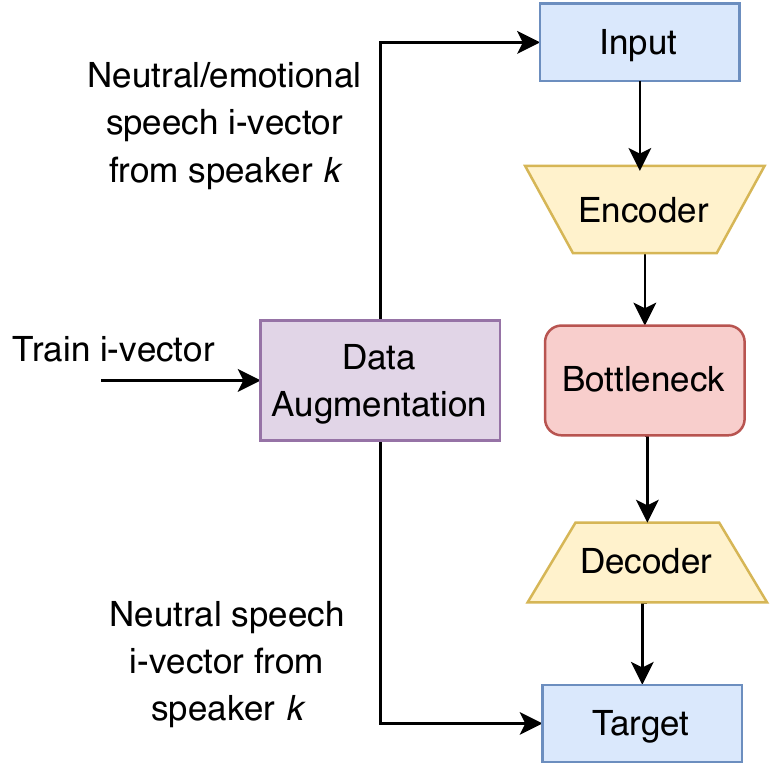}
%\includegraphics[height=100mm,width=115mm]{train3.jpg}
%\vspace{-1mm}
\caption{ \label{training} Block diagram representing training of emotion invariant extractor.}
%\vspace{-0.4cm}
\end{figure}

%An autoencoder model is trained to map the i-vectors computed from the neutral/  emotional speech to i-vector of neutral speech. The method of training the model for the speaker identification task is described as follows.

%The difference in emotion between train and test speech produces a mismatch that has been found to reduce the performance for speaker identification as observed from the literature review from Introduction section. Generally, speaker models are recorded in neutral speech and the test speech in practical settings can contain any emotional state of speaker. In this regard, we plan to learn an extractor that maps the i-vector based speaker representation of different emotions to neutral speech category. The embedding produced from such extractor is expected to work well for mismatch emotion scenarios. We refer the speaker embedding thus produced as emotion invariant speaker embedding.

In general, speaker enrollments can be done with neutral speech. However, test speech can contain any emotional state in practical scenarios. Therefore, we plan to learn an extractor that maps the i-vector based speaker embeddings of different emotions to that of neutral state. We refer the speaker embedding thus produced as emotion invariant speaker embedding.

% \begin{figure*}[t!]
% \centering
% \includegraphics[width=\textwidth]{spk_id3.jpg}
% %\includegraphics[height=52.5mm,width=180mm]{spk_id2.jpg}
% %\vspace{2mm}
% \caption{ \label{inference}
% Block diagram showing proposed emotion invariant speaker embedding based system for speaker identification.}
% \vspace{-0.2cm}
% \end{figure*}

Figure \ref{training} shows the training process of the emotion invariant extractor. A deep neural network with three hidden layers, namely, encoder, bottleneck and decoder are used in the process. The i-vectors of neutral or any emotional speech from a particular speaker is used as input and that of neutral speech belonging to the same speaker is used as the target to train the network. The input dimension equals to the i-vector dimension. A data augmentation process is used to increase the number of training i-vectors of different emotions. The details of this process is described in Section~\ref{seciii_iii}. %We used two parameter sets ($P_{set1}$ and $P_{set2}$) for experimentation, one with the linear activation function and the other with sigmoid activation function.
%In the first set ($P_{set1}$), the encoder , bottleneck and decoder layer have 32, 16 and 32 hidden neurons, respectively. In the second set ($P_{set2}$), the encoder , bottleneck and decoder layer have 64, 32 and 64 hidden neurons, respectively. The final layer in both the sets has 150 output units.

We used linear activation function at the final layer as the i-vectors can have any real values. The ReLU activation function is used for all the layers except the final layer. Mean squared error is used as the loss function and Adam optimizer is used for optimizing the network \cite{kingma2014adam}. The model is trained for 20 epochs with batch size of 256. The various parameters used in the network are summarized in Table \ref{table1_param}. We note that $\beta_1$ and $\beta_2$ are parameters of Adam optimizer used in the network. 

%Next, we discuss the speaker identification framework using this emotion invariant extractor. 

\begin{table}[t!]
\centering
\caption{Details of emotion invariant extractor.}
%\vspace{-2mm}
\label{table1_param}
%\resizebox{6cm}{!}{
\begin{tabular}{|c|c|}
\hline
%\multirow{2}{*}{\bf{Parameters}} & \multicolumn{2}{|c|}{\textbf{Value Sets}} \\
%\cline{2-3}
%&\textbf{$P_{set1}$}&\textbf{$P_{set2}$}\\
{\bf{Parameters}} & {\bf{Value Sets}}\\
\hline
\hline
Input & 150\\
\hline
Encoder layer&64\\
\hline
Activation  &ReLU \\
\hline
Bottleneck layer&32\\
\hline
Activation & ReLU\\
\hline
Decoder layer &64\\
\hline
Activation  & ReLU\\
\hline
Output Layer & 150\\
\hline
Activation  & Linear\\
\hline
\hline
Loss Function & {Mean Squared Error}\\
\hline
Optimizer & {Adam}\\
\hline
Learning Rate & {0.001}\\
\hline
$\beta_1$ & {0.9}\\
\hline
$\beta_2$ & {0.999}\\
\hline
\# Epochs & {20}\\
\hline
\end{tabular}
%}
%\vspace{-3mm}
\end{table}

\subsection{Speaker Identification with Emotional Speech}

%Figure \ref{inference} shows the speaker identification framework with the proposed emotion invariant speaker embedding.

%The i-vector based speaker models for the studies are created with neutral speech corresponding to each speaker. During testing,

For a given speech irrespective of any emotion, its i-vector is obtained, followed by extraction of corresponding emotion invariant speaker embedding. We use this process with two different ways in terms of considering the speaker models. The first one refers to EINV-Test that extracts emotion invariant speaker embeddings only for the test data and uses the raw train i-vectors as speaker models. On the other hand, the second one is referred to as EINV-Pair that extracts emotion invariant speaker embeddings for both train and test data. It is noted that the baseline system follows the standard i-vector pipeline without any emotion invariant extractor.

%The resultant embeddings thus obtained are then used to identify the closest matched speaker model. We consider two different frameworks with this, where the the first one only creates emotion invarianet speaker embeddings for the test data. On the other hand the second fraemwork, creates for both train and test data.

%It is noted that the baseline system in this work follows the same pipeline without the emotion invariant extractor.  

%Given a speech, first its features are extracted, followed by computation of corresponding i-vectors. Then irrespective of any emotion class at the input signal its emotion invariant embedding is derived by using the extractor described in previous subsection. Finally, the test i-vector is compared to the speaker models to identify the closest matched speaker. We note that the baseline system for our studies has similar pipeline without the emotion invariant speaker embedding extractor. 

%Next, we discuss the details of the experiments conducted in this work. 

%process using i-vector transformation. First MFCC is extracted from the neutral or emotional speech input and then i-vector is calculated from the MFCC feature vector. The i-vector is then transformed using the trained autoencoder model. The transformed i-vector is compared with the training i-vectors using the cosine kernel metric to predict the speaker.

%\section{Challenges in Identifying Speakers with Different Emotional States}
%\label{secii}

\section{Experiments}
\label{seciii}

In this section, we discuss the details of the corpus, feature extraction, experimental setup for baseline and proposed framework in the following subsections.
%speaker identification with emotional speech. The following subsections mention their description.  

\subsection{Database}

The IEMOCAP database is considered for the studies in this work. It is an acted speech database from 10 different speakers comprising of 5 male and 5 female. The database is segmented and each segment is labeled with one of the emotion classes. We consider four emotion classes, namely, neutral, happiness, anger and sadness for the study due to availability of sufficient amount of the data from those emotions in the corpus. %The speech data from those four emotion classes are considered for the study.
First two minutes of each emotion class are used for the training and the rest for testing. The speaker models are created using the 2 minutes of training data, whereas the test data is divided into many non-overlapping 30 seconds segments and each of those segments are used for testing. A summary of the training and testing examples is shown in Table \ref{table2}. %Each speaker has four training examples from four different emotions to have four different models for the studies.
Further, we note that the Switchboard Corpus-II is used for learning the background models for the i-vector system. 

\begin{table}[t!]
\centering
\caption{Summary of the corpus with different emotion classes namely, neutral (N), happiness (H), anger (A) and sadness (S).}
%Training segments are of 2 minutes duration and testing segments are of 30 seconds duration.}
%\vspace{-2mm}
\label{table2}
%\resizebox{8.1cm}{!}{
\begin{tabular}{|c||c|c|c|c||c|c|c|c|}
\hline
\textbf{Speaker} & \multicolumn{4}{|c||}{\textbf{\# Train Utterances}}& \multicolumn{4}{|c|}{\textbf{\# Test Utterances}} \\
\cline{2-9}
\textbf{Label}&\textbf{N} & \textbf{H} & \textbf{A} & \textbf{S} & \textbf{N} & \textbf{H} & \textbf{A} & \textbf{S}\\
\hline
\hline
01F&1&1&1&1&20&10&25&16\\
\hline
01M&1&1&1&1&29&10&15&22\\
\hline
02F&1&1&1&1&20&10&10&18\\
\hline
02M&1&1&1&1&35&7&10&22\\
\hline
03F&1&1&1&1&16&11&16&25\\
\hline
03M&1&1&1&1&26&9&22&23\\
\hline
04F&1&1&1&1&9&6&30&13\\
\hline
04M&1&1&1&1&23&8&16&19\\
\hline
05F&1&1&1&1&25&10&11&22\\
\hline
05M&1&1&1&1&24&11&14&23\\
\hline
% \hline
% Total & 10 & 10 & 10 & 10 & 277 & 92 & 167 & 193\\
% \hline
\end{tabular}
%}
%\vspace{-2mm}
\end{table}

%\subsection{I-vector data augmentation}

\subsection{Feature Extraction}

%The speech utterances of the IEMOCAP corpus are first downsampled to 8 kHz as the background data from Switchboard Corpus II are in 8 kHz sampling rate. %We used widely popular MFCC features for our studies~\cite{Davis1980}.
The speech signals are short-term processed with 20 ms frame with a shift of 10 ms to extract 39-dimensional (13-base+13-$\Delta$+13-$\Delta\Delta$) MFCC features~\cite{Davis1980}. An energy based voice activity detection is performed to consider the regions with sufficient voice activity~\cite{energy_vad}. The features of those regions are then normalized by cepstral mean and variance normalization to fit to zero and unit variance~\cite{Furui1981}. 
%We note that the background data from Switchboard corpus are processed similarly to obtain corresponding features. 

\subsection{Experimental Setup: Baseline i-vector}

The background data features from Switchboard-II corpus are used to train the background models for i-vector system. Equal amount of male and female data is chosen to train a 1024 component gender-independent UBM. The zeroth and the first order statistics of the background data are then extracted to learn the T-matrix of 400 columns. The features belonging to IEMOCAP corpus are then used to obtain the sufficient statistics followed by transformation using the T-matrix to derive the 400-dimensional i-vectors. Further, we have learned 150-dimensional LDA and full dimensional WCCN using the background data i-vectors. The LDA and WCCN are then applied on top of the i-vectors for channel/session compensation, which reduces their dimension to 150. For the studies with baseline system, we used the channel/session compensated i-vectors of train and test data to perform a cosine similarity scoring for identifying a speaker with the highest similarity.

\subsection{Experimental Setup: Emotion Invariant Extractor}
\label{seciii_iii}

In order to create the emotion invariant extractor, sufficient amount of training data with emotional speech is required. The data should contain i-vectors of neutral or emotional speech at the input layer and i-vectors of neutral speech of the same speaker at the output layer of the network. Since we have only 2 minutes of training data per speaker for each emotion, data augmentation is performed to enhance the training.

\begin{figure}[t!]
\centering
%\vspace{2mm}
\includegraphics[width=0.48\textwidth]{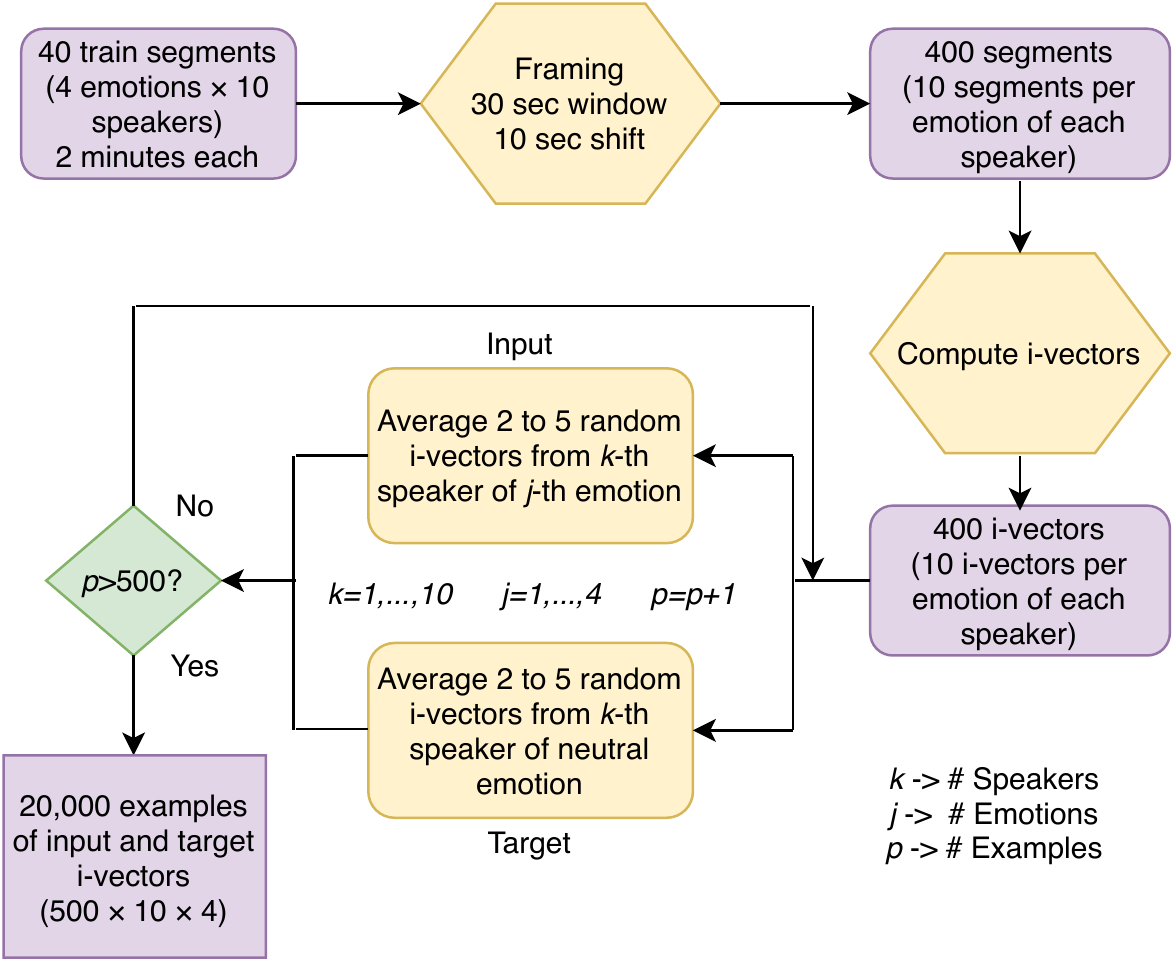}
%\includegraphics[height=52.5mm,width=180mm]{spk_id2.jpg}
%\vspace{3mm}
\caption{ \label{data_aug}
Block diagram showing the data augmentation procedure used in training of emotion invariant extractor.}
%\vspace{-2mm}
\end{figure}

First, the 2 minutes data is split into many overlapping 30 seconds segments at every 10 second interval. This resulted in 10 small segments per speaker for each emotion class. Total number of such training segments becomes 400 with 10 speakers and 4 emotion classes. The i-vectors for all such segments are extracted as described in the previous subsection. Next, data augmentation is performed by averaging i-vectors of two to five segments from the same speaker with different combinations. This augmentation is performed for producing both input and target i-vectors. In this way, 20,000 input and target i-vectors are generated for training. The details of the data augmentation process is shown in Figure~\ref{data_aug}. We note that the augmented i-vectors are divided into training and validation sets for learning the emotion invariant extractor; 80\% of the data is used for training and rest 20\% is used for validation. 
%However, due to high correlation in the data, the convergence curves of the training and the validation data are found to be almost similar.

The emotion invariant extractor thus learned, considers 150-dimensional input i-vectors from either neutral or any other emotional speech to generate corresponding emotion invariant speaker embedding. Once the emotion invariant speaker embedding is obtained, the remaining stages for speaker identification studies follows the pipeline described earlier. In this work, identification accuracy (in \%) is used as metric to report the results of various frameworks. Next, we discuss the results and analysis for the studies. 

% \section{Auto-encoder based i-vector Transformation}
%\section{Emotion Invariant Speaker Embedding}
%\label{seciv}

\begin{table}[t!]
%\vspace{-2mm}
\centering
\caption{Speaker identification performance in accuracy (\%) for different train and test conditions using emotion classes, namely, neutral (N), happiness (H), anger (A) and sadness (S).}
%The Cmb1 and Cmb2 refer to conditions when training data from all four emotions are used in two different ways, whose details are mentioned in text.}
%\vspace{-2mm}
\label{table3}
%\resizebox{8.1cm}{!}{
\begin{tabular}{|c||c|c|c|c|}
\hline
{\textbf{Test}} & \multicolumn{4}{|c|}{\textbf{Train Emotion}} \\
\cline{2-5}
\textbf{Emotion}&{\textbf N}&{\textbf H}&{\textbf A}&{\textbf S} \\
\hline\hline
 N & {\bf 92.9} & 87.2 & 83.7& 89.4\\
\hline
H & 82.6 & {\bf 85.8} & 70.6 & 73.9\\
\hline
A & 69.2 & 62.1 & {\bf 85.2} & 54.4\\
\hline
S & {\bf 82.2} & 80.2 & 69.9 & 81.7\\
\hline
\hline
Average & {\bf 81.7} & 78.8 & 77.3 & 74.8\\
\hline
\end{tabular}
%}
%\vspace{-1mm}
\end{table}

\section{Results and Analysis}
\label{seciv}

We first evaluate the performance of baseline i-vector system for identifying speakers with emotional speech. Table \ref{table3} reports its performance for different train and test conditions using the four emotion classes. We observe that the average accuracy is high when neutral emotion is used for speaker modeling. In addition, the accuracy is maximum when neutral speech considered in both training as well as testing and a comparatively a higher performance is obtained when the train and test emotions have a match. However, for sadness emotion, the performance is maximum when the training emotion is neutral. As a summary, we find that the overall performance is comparatively better when the training data contains neutral emotion speech from the speakers. This clearly supports the motivation behind the work, i.e., to map the i-vector based speaker representation of different emotions to that of the neutral speech to derive emotion invariant speaker embeddings.

We now pay our attention to the studies with proposed emotion invariant speaker embeddings, where the i-vectors of emotional speech are converted to that of the neutral speech. At this stage, we also build a contrast system for comparing to our proposed systems EINV-Test and EINV-Pair. The contrast system considers speaker models that contain information of multiple emotions, thereby showing scope for having a better match to different emotions during testing. In our case, we consider i-vector averaging of different emotions to derive the speaker models for the contrast system. It is to be noted that we also perform averaging of train i-vectors belonging to each speaker with our proposed EINV-Test and EINV-Pair frameworks.

%In Table \ref{table3}, performance is also evaluated by considering training data from all four emotion classes in two different ways. The Cmb1 and Cmb2 denote these two methods in the table. In Cmb1, the training i-vectors from all four emotion classes of each speaker are averaged and then compared with the testing i-vector. We observe improvement for the neutral and sadness classes with this speaker representation. On the other hand, in Cmb2, testing i-vector is compared with training i-vectors from all the emotions of every speaker and decision is made based on the maximum similarity in the cosine kernel output. With this combination approach, we find absolute improvements of 5.4\%, 3.3\% and 2\% for neutral, happiness and sadness emotions, respectively. However, 4\% decrement is also observed for anger emotion compared to Cmb1. In this study, these two methods, Cmb1 and Cmb2 are used as benchmark baseline systems. We also report the unweighted accuracy for all the cases discussed in Table~\ref{table3}.

Table~\ref{table4} shows the performance comparison of the contrast system and proposed emotion invariant speaker embedding system with two different frameworks EINV-Test and EINV-Pair. First, on observing the performance of the contrast system based on average train i-vectors with different emotions, we find that it performs better than our previous study shown in Table~\ref{table3}. The improvements are prominent for neutral and sadness emotions during testing. Further, the best average accuracy of 81.7\% obtained in Table~\ref{table3} with neutral emotion based speaker model improves to 87.9\%, when averaging is performed on the train i-vectors with different emotions. This shows that the contrast system itself has potential to reduce the emotion mismatch by capturing information from different emotions.
%in the speaker models. 

\begin{table}[t!]
\centering
%\vspace{-2mm}
\caption{Performance comparison in accuracy (\%) for speaker identification using different frameworks with speech from four emotion classes, namely, neutral (N), happiness (H), anger (A) and sadness (S).}
%The results are obtained using parameter set $P_{set1}$ with linear activation function.}
%\vspace{-2mm}
\label{table4}
%\resizebox{8.1cm}{!}{
\begin{tabular}{|c||c||c|c|}
\hline
%\multirow{2}{*}
{\textbf{Test}}&\multicolumn{3}{|c|}{\textbf{Framework}}\\
\cline{2-4}
\cline{2-4}
%&\multicolumn{2}{|c||}{\textbf{Baseline}}&\multicolumn{2}{|c|}{\textbf{Proposed Emotion Invariant Speaker  Embedding}}\\
%\cline{2-4}
%&\multirow{2}{*}{\textbf{Avg. i-vector}}&\multirow{2}{*}{\textbf{Best Score}}&\multicolumn{2}{|c|}{\textbf{Avg. i-vector}}\\ 
%\cline{4-5}
% &&&\textbf{Train-embed}&\textbf{Train-raw}\\
%&\textbf{(Cmb1)}&\textbf{(Cmb2)}&\textbf{(Prop1)}&\textbf{(Prop2)}\\
{\textbf{Emotion}}&{\textbf{Avg. i-vector}} & {\textbf{EINV-Test}}&{\textbf{EINV-Pair}}\\
\hline\hline
N & 93.8 & 91.6 & 92.5 \\
\hline
H & 85.8 & 91.3 & 91.3\\
\hline
A & 85.2 & 88.1 & 89.3\\
\hline
S & 86.6 & 89.6 & 89.1\\
\hline
\hline
Average & 87.9 & {\bf90.1} & {\bf90.5}\\
\hline
\end{tabular}%}
%\vspace{-2mm}
\end{table}

% \begin{figure}[t!]
% \centering
% \includegraphics[width=0.68\textwidth]{hist.jpg}
% %\includegraphics[height=52.5mm,width=180mm]{spk_id2.jpg}
% %\vspace{2mm}
% \caption{ \label{hist_plot}
% Histogram of (a) raw i-vector values (b) scaled (halfed) i-vector values}
% \vspace{-2mm}
% \end{figure}

% \begin{figure}[t!]
% \centering
% %\includegraphics[width=\textwidth]{rel_impr.png}
% \includegraphics[width=0.5\textwidth]{activations2.jpg}
% %\vspace{1mm}
% \caption{ \label{activations}
% Performance comparison with different activation functions}
% \vspace{-2mm}
% \end{figure}

% \begin{figure}[t!]
% \centering
% %\includegraphics[width=\textwidth]{rel_impr.png}
% \includegraphics[width=0.45\textwidth]{rel_impr2.png}
% %\vspace{1mm}
% \caption{ \label{rel_impr}
% Histogram showing relative improvements in identification accuracy of different proposed methods.}
% \vspace{-2mm}
% \end{figure}

We now focus on the performance comparison among the contrast system with average i-vectors of emotional speech and the proposed frameworks with emotion invariant speaker embeddings in Table~\ref{table4}. It is observed that the performance of speaker identification for happiness, anger and sadness emotion consistently improves with both EINV-Test and EINV-Pair frameworks. Further, we find that speaker identification with happiness emotion is benefited maximum, whereas the sadness and anger emotions are more or less equally impacted for improvements. It is also to be noted that there is a decrease in performance for the case of neutral emotion, which may be due to the unnecessary transformation made on the i-vectors when they are already obtained from neutral speech.

Further, on comparing the average accuracy, we find both EINV-Test and EINV-Pair outperform the contrast i-vector averaging of multiple emotion based system, showing the effectiveness of proposed emotion invariant speaker embeddings. An average accuracy of 90.5\% is obtained with EINV-Pair framework, which has an absolute improvement of 2.6\% over the average i-vector based approach with different emotions. We also note the improvements with EINV-Pair are slightly higher than that of EINV-Test as both train and test i-vectors are transformed to emotion invariant space, thereby further reduces the emotion mismatch.

\begin{figure}[t!]
\centering
\includegraphics[width=0.47\textwidth]{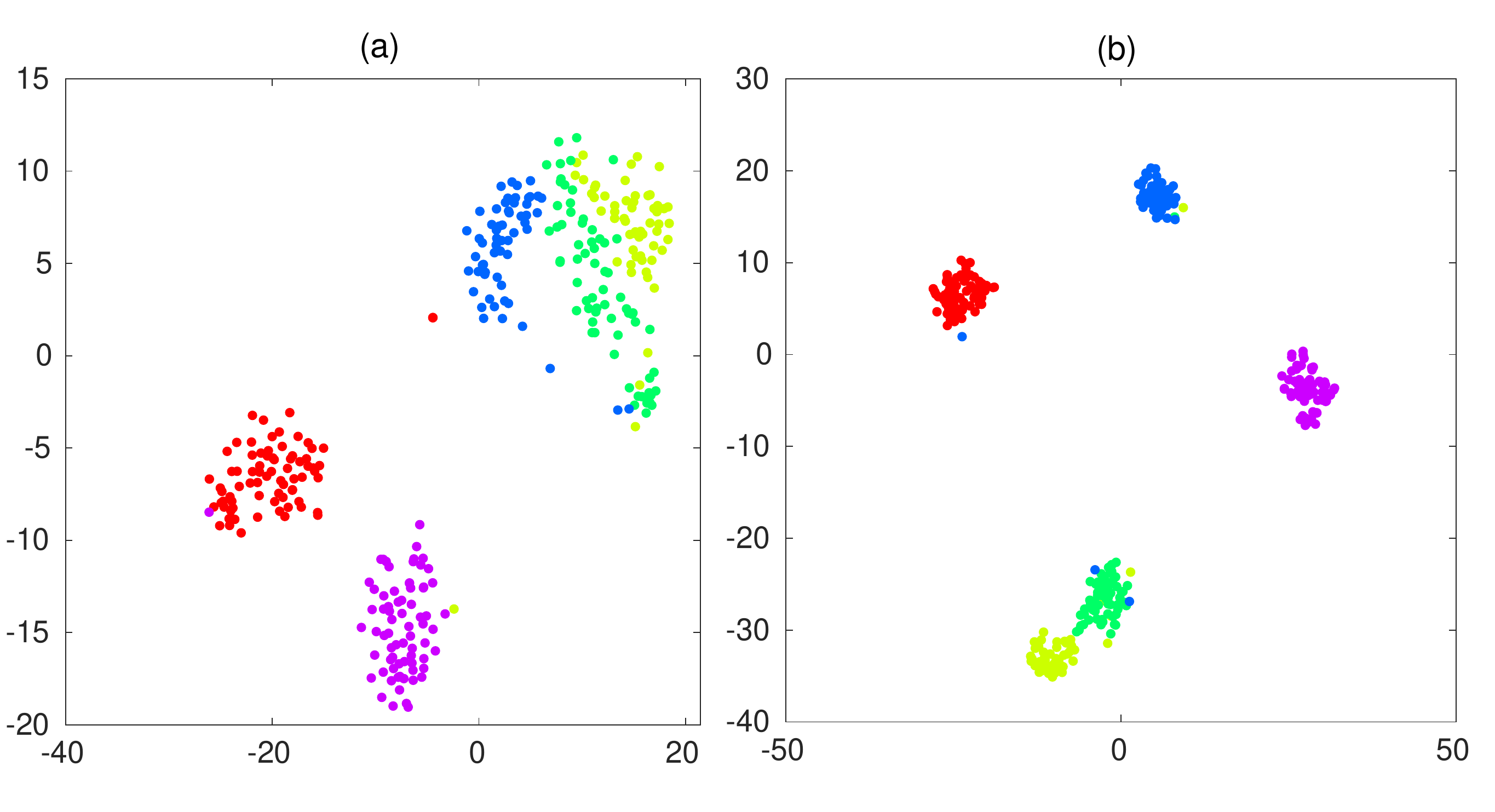}
%\includegraphics[height=52.5mm,width=180mm]{spk_id2.jpg}
%\vspace{2mm}
\caption{ \label{tSNE_plot}
t-SNE visualizations for 5 different speakers: (a) baseline i-vectors (b) emotion invariant speaker embeddings. The five different colors represent speaker labels.}
%\vspace{-2mm}
\end{figure}

We now plot the t-SNE visualizations to observe the effect of proposed emotion invariant speaker embeddings on i-vectors from different speakers with emotional speech~\cite{Maaten2008Visualizing}. Figure~\ref{tSNE_plot} shows the t-SNE plots for randomly chosen 5 speakers from the database for this analysis. The five different colors represent i-vectors with four different emotions from those different 5 speakers. We find that the emotion invariant speaker embeddings are more distinguishable as speaker clusters and separable compared to the original i-vector based speaker representation. This further strengthens the effectiveness of the proposed emotion invariant speaker embedding for identifying speakers with emotional speech.

In this work, we have studied speaker identification with emotional speech using a relatively smaller database due to unavailability of a larger corpus for such studies. The future work will focus on extending the work on a relatively larger database with more number of emotion classes as well as speakers to demonstrate its significance for real-world scenario.

%In addition, we intend to explore emotion classification followed by use to this framework to avoid the unnecessary transformation of neutral i-vectors to the emotion invariant embedding. 

%The future work

\section{Conclusions}
\label{conc}
This work attempts to improve speaker identification with emotional speech from the view of practical systems. Four different emotions, namely, neutral, anger, happiness and sadness from IEMOCAP database are considered for the study. A baseline and another contrast system are built using training data from the four emotions and their average models with different emotions using i-vector modeling. A novel method is then proposed to transform the i-vectors containing speaker-specific information into an emotion invariant space in terms of emotion invariant speaker embedding. This proposed representation gives an absolute improvement of 2.6\% in accuracy over the average
speaker model with different emotions based system. In addition, we observe significant improvements for happiness, anger and sadness emotion classes, which is maximum for the happiness emotion.

\section{Acknowledgements}

The work of the second author is supported by Programmatic Grant No. A1687b0033 from the Singapore Government's Research, Innovation and Enterprise 2020 plan (Advanced Manufacturing and Engineering domain), and Human-Robot Interaction Phase 1 (Grant No. 192 25 00054) by the National Research Foundation, Prime Minister's Office, Singapore under the National Robotics Programme.

%\newpage
\balance
\bibliographystyle{IEEEtran}

\bibliography{MyReferences_new}

\end{document}